\documentclass[a4paper,fleqn,usenatbib]{mnras}

\usepackage[T1]{fontenc}
\usepackage{ae,aecompl}

\usepackage{graphicx}	% Including figure files
\usepackage{amsmath}	% Advanced maths commands
\usepackage{amssymb}	% Extra maths symbols
\usepackage{multirow}

\newcommand{\swift}{{\it Swift}}
\newcommand{\xmm}{{\it XMM-Newton}}

\title[Is there a UV/X-ray connection in IRAS~13224--3809?]{Is there a UV/X-ray connection in IRAS~13224--3809?}

\author[D. J. K. Buisson et al.]{D. J. K. Buisson$^{1}$,\thanks{Email: djkb2@ast.cam.ac.uk}
	A. M. Lohfink$^{1,2}$,
	W. N. Alston$^{1}$,
	E. M. Cackett$^{4}$,
\newauthor	C.-Y. Chiang$^{4}$,
	T. Dauser$^{5}$,
	B. De Marco$^{3}$,
	A. C. Fabian$^{1}$,
	L. C. Gallo$^{6}$,
\newauthor	J. A. Garc\'ia$^{7,5,8}$,
	J. Jiang$^{1}$,
	E. Kara$^{9}$,
	M. J. Middleton$^{10}$,
	G. Miniutti$^{11}$,
\newauthor	M. L. Parker$^{1}$,
	C. Pinto$^{1}$,
	P. Uttley$^{12}$,
	D. J. Walton$^{1}$ and
	D. R. Wilkins$^{13}$  \\
  $^{1}$Institute of Astronomy, Madingley Road, Cambridge, CB3 0HA\\
  $^{2}$Montana State University, Bozeman, 59717-3840, MT, USA\\
  $^{3}$Nicolaus Copernicus Astronomical Center, Polish Academy of Sciences, Bartycka 18, PL-00-716 Warsaw, Poland\\
  $^{4}$Department of Physics \& Astronomy, Wayne State University, 666 W. Hancock St, Detroit, MI 48201, USA\\
  $^{5}$Dr. Karl Remeis-Observatory and Erlangen Centre for Astroparticle
Physics, Sternwartstr. 7, 96049 Bamberg, Germany\\
  $^{6}$Department of Astronomy and Physics, Saint Mary's University, 923 Robie Street, Halifax, NS B3H 3C3, Canada\\
  $^{7}$Cahill Center for Astronomy and Astrophysics, California Institute of Technology, Pasadena, CA 91125\\
  $^{8}$Harvard-Smithsonian Center for Astrophysics, 60 Garden St, Cambridge, MA 02138, USA\\
  $^{9}$Department of Astronomy, University of Maryland, College Park, MD 20742-2421, USA\\
  $^{10}$Department of Physics and Astronomy, University of Southampton, Highfield, Southampton, SO17 1BJ\\
  $^{11}$Centro de Astrobiolog\'ia (CSIC--INTA), Dep. de Astrof\'isica; ESAC, PO Box 78, Villanueva de la Ca\~nada, E-28691 Madrid, Spain\\
  $^{12}$Anton Pannekoek Institute, University of Amsterdam, Science Park 904, 1098 XH Amsterdam, The Netherlands\\
  $^{13}$Kavli Institute for Particle Astrophysics and Cosmology, Stanford University, 452 Lomita Mall, Stanford, CA 94305, USA
}

\date{Received 2017 December 18; in original form 2017 October 2}

\pubyear{2017}

\begin{document}
\label{firstpage}
\pagerange{\pageref{firstpage}--\pageref{lastpage}}
\maketitle
\begin{abstract}
We present results from the optical, ultraviolet and X-ray monitoring of the NLS1 galaxy IRAS~13224--3809 taken with \swift\ and \xmm\ during 2016.
IRAS~13224--3809 is the most variable bright AGN in the X-ray sky and shows strong X-ray reflection, implying that the X-rays strongly illuminate the inner disc. Therefore, it is a good candidate to study the relationship between coronal X-ray and disc UV emission.
However, we find no correlation between the X-ray and UV flux over the available $\sim40$\,day monitoring, despite the presence of strong X-ray variability and the variable part of the UV spectrum being consistent with irradiation of a standard thin disc.
This means either that the X-ray flux which irradiates the UV emitting outer disc does not correlate with the X-ray flux in our line of sight and/or that another process drives the majority of the UV variability.
The former case may be due to changes in coronal geometry, absorption or scattering between the corona and the disc.
\end{abstract}


\section{Introduction}

AGN are the most luminous persistent point sources in the Sky in the optical to X-ray bands. They have a significant impact on galaxy evolution and are therefore of great interest for study.
Since AGN are unresolved with current instruments in the X-ray band, their structure must be inferred from properties of their spectra or the variability of their emission.

A significant fraction of their bolometric luminosity is emitted in the X-ray band from a small region known as the corona \citep{haardt93,merloni03}. Microlensing \citep{dai10, chartas12} and timing \citep{demarco11,demarco13,reis13,kara14,kara16} results show that this is often smaller than $10r_g$ in size.
Much of the X-ray power from the corona is directed towards the accretion disc, as seen in reflection features in the X-ray spectrum \citep{tanaka95,fabian10}.

Additional evidence that the X-ray emission affects the disc is that variations in X-ray and optical fluxes are often seen to be correlated \citep[e.g.][]{alston13,shappee14,edelson15,fausnaugh16,buisson17,gliozzi17,lobban17}. Where the optical emission lags the X-rays, this is often interpreted as heating of the disc by the additional X-ray flux directed towards the disc \citep{lightman88}. In some cases \citep[e.g.][]{troyer16,edelson17}, the lags are longer than predicted for a standard thin disc \citep{shakura73} and the X-ray lightcurve does not always match the inferred driving lightcurve \citep{starkey17}. This may be explained by a larger disc or an additional stage of reprocessing \citep{gardner17,edelson17}. There is also now good evidence that diffuse continuum emission from the broad line region can also contribute significantly to the lags, which needs to be accounted for \citep{cackett17,mchardy17}. Sometimes, the optical emission is found to lead the X-ray emission \citep{arevalo05}, which is interpreted as Compton upscattering of the optical photons to X-rays \citep{haardt91} or the propagation of fluctuation inwards through the disc \citep{lyubarskii97, arevalo06}.
However, sometimes no correlation is found \citep[e.g.][]{robertson15}.
Continued study of optical to X-ray variability in more sources has the potential to provide more information on why correlations are seen only in some sources.

The narrow line Seyfert 1 (NLS1) galaxy IRAS~13224--3809 ($z=0.066$, $M_{\rm BH}=10^{6}-10^{7}\,{\rm M_\odot}$, \citealt{zhou05}) is the most variable AGN in X-rays, often showing changes in X-ray flux by a factor of 50 on timescales of less than one hour \citep{boller97,dewangan02,fabian13}.
Its X-ray spectrum shows a soft continuum with strong relativistic reflection and soft excess \citep[][Jiang et al. submitted]{ponti10,fabian13,chiang15}.
The soft X-ray continuum suggests that IRAS~13224--3809 is accreting at a high Eddington fraction ($\dot{m}\simeq0.7$ using the relation from \citealt{shemmer08}).
It shows little X-ray obscuration, although the recent \xmm\ observations have allowed the detection of an Ultra-Fast Outflow (UFO) which is observed only at low X-ray flux \citep{parker17nat,parker17var}.
Previous studies show that IRAS~13224--3809 has little absorption in the UV and that the \ion{C}{IV} emission line is asymmetric and blueshifted \citep{leighly04a,leighly04b}, which may indicate an outflow out of the line of sight.

The strong X-ray variability and reflection suggest strong variable heating of the disc, so IRAS~13224--3809 is an ideal candidate to study UV/X-ray relations.
The source is a member of the sample studied in \citet{buisson17} to find UV/X-ray relations. This work found a marginally significant ($2\sigma$ confidence) lag of \textit{UM2}-band ($\sim2170$\,\AA) behind X-ray emission, suggesting that X-ray reprocessing may occur in this source. 
Here, we present the results from the Optical Monitor of the recent 1.5\,Ms \xmm\ observing campaign of IRAS~13224--3809, along with associated \swift\ monitoring (50\,ks XRT exposure over the period 7$^{\rm th}$ July to 14$^{\rm th}$ August).

The increase in data now available allows us to study more of its properties. The additional \swift\ monitoring allows us to measure the optical/UV variable spectrum and the extensive \xmm\ coverage provides constraints on the short timescale UV/X-ray relation.

\section{Observations and Data Reduction}
\label{section_datareduction}

\subsection{\xmm}

\begin{figure}
\includegraphics[width=\linewidth, clip, trim={20cm 2cm 15cm 4cm}]{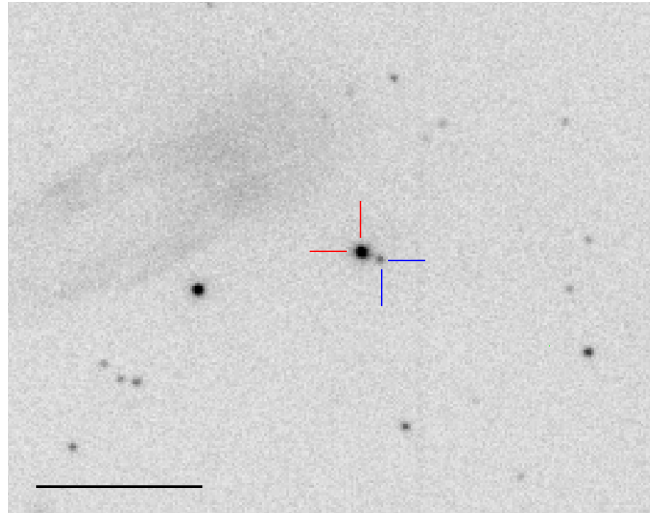}
\caption{Image from \xmm-OM showing IRAS~13224--3809 (red, left) and nearby secondary source (blue, right). The scalebar indicates 1\,arcmin.}
\label{fig:omimage}
\end{figure}

\begin{figure*}
\includegraphics[width=\linewidth]{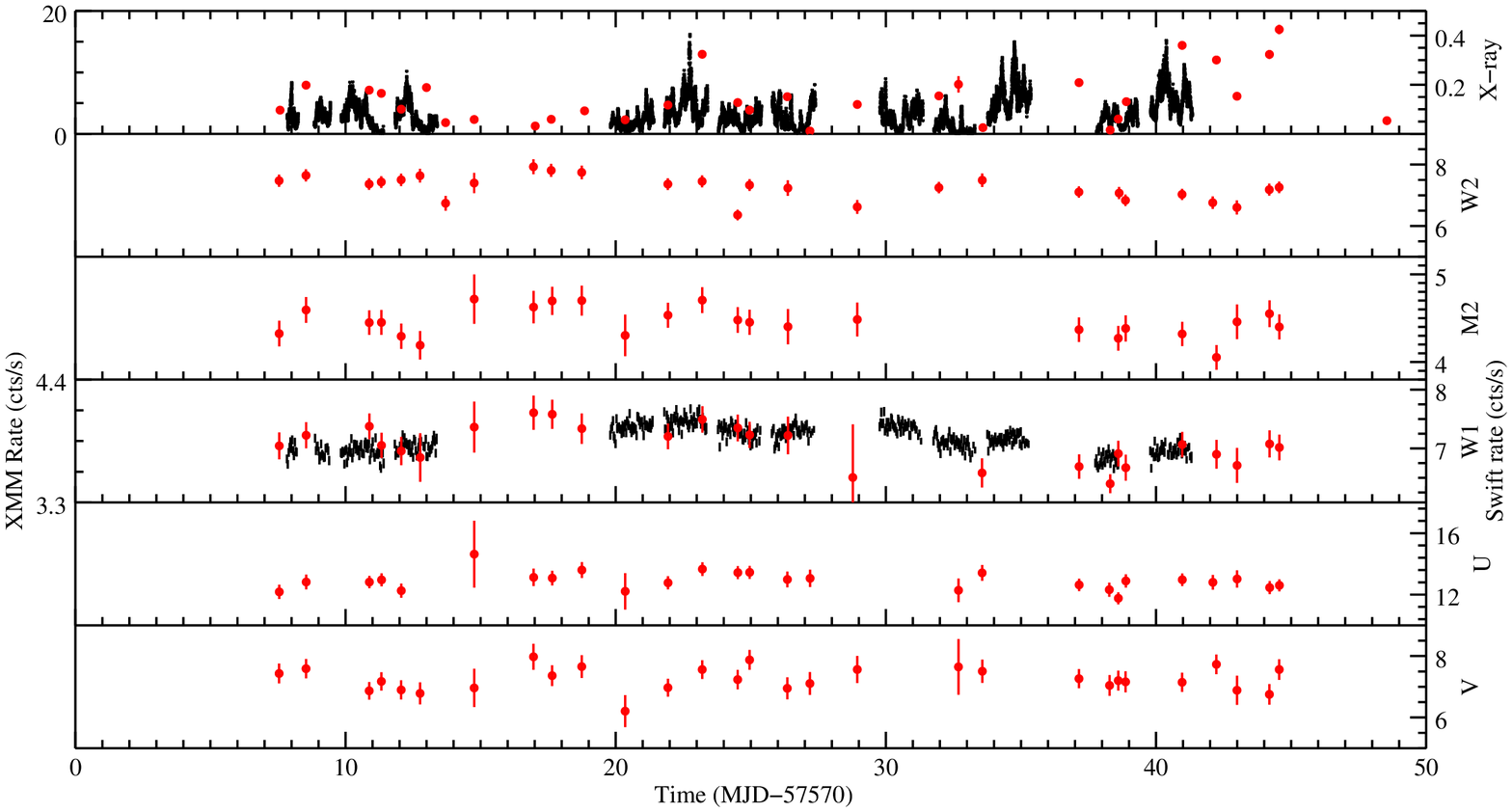}
\includegraphics[width=\linewidth]{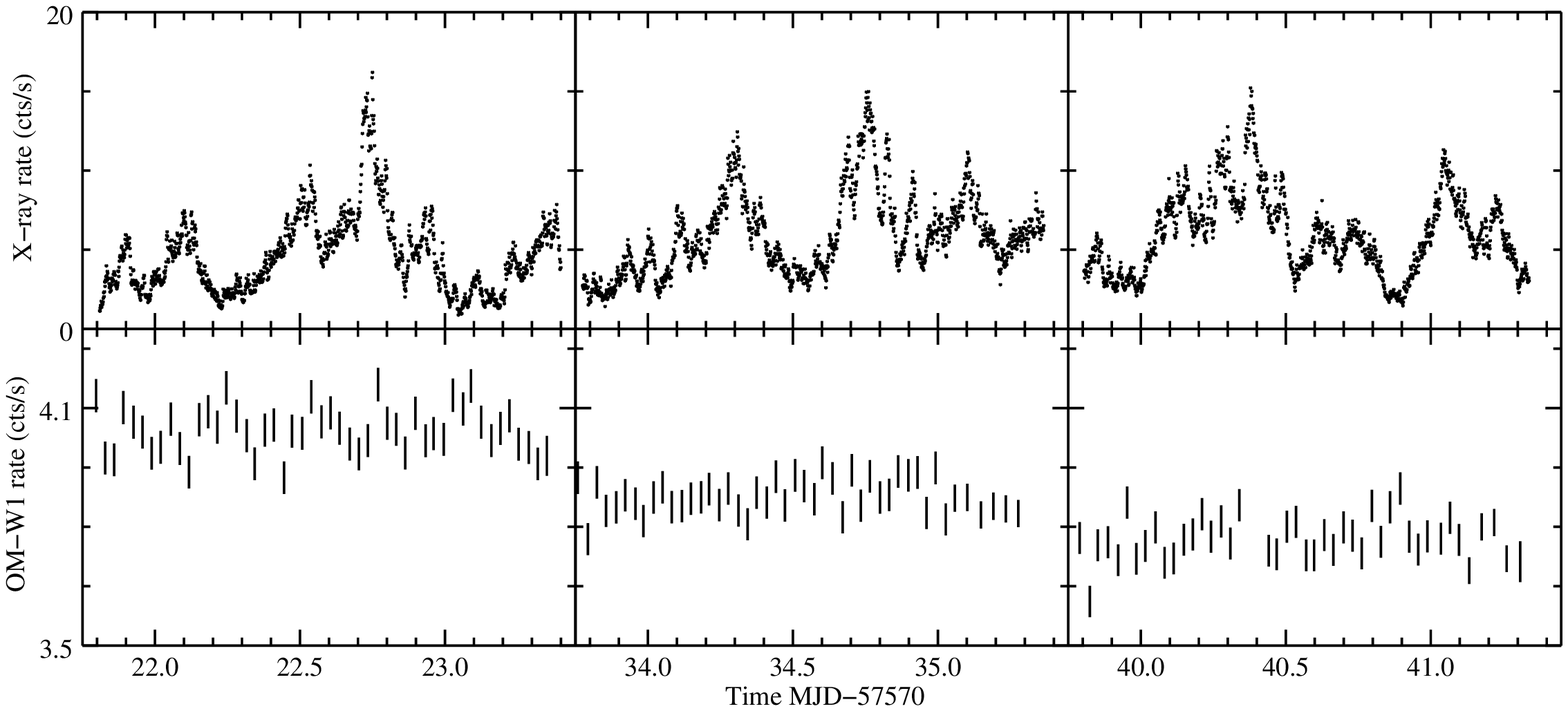}
\caption{Lightcurves of IRAS~13224--3809 from \xmm\ (black) and \swift\ (red). Upper panels show, from top to bottom: X-rays ($0.3-10$\,keV), {\it W2}-band, {\it M2}-band, {\it W1}-band, {\it U}-band, {\it V}-band. Note that the W1 filters of \swift\ and \xmm, although plotted in the same panel, are not identical.
Lower panels show detail of the X-ray and UV lightcurves of the three \xmm\ orbits with the strongest X-ray peaks. There is no apparent response of the UV emission to the X-ray peaks.}
\label{fig:lightcurves}
\end{figure*}

We use \xmm\ \citep{jansen01} data from the recent very large programme (P.I. Fabian) dedicated to monitoring IRAS~13224--3809, with observations from July to August 2016.
Here, we consider X-ray lightcurves from the EPIC-pn \citep{struder01} instrument and ultraviolet lightcurves from the Optical Monitor (OM, \citealt{mason01}). To provide continuous coverage, the OM observations were taken in the {\it W1}-band throughout and used a typical frametime of 2700\,s.

The pn data were reduced using the standard task \textsc{epchain}, using a 50\,arcsec circular source region and an annular background region comprising radii from 60--90\,arcsec.
Data were taken in Large Window mode, leading to mild pileup in the brightest X-ray states.
While this may affect the detail of the X-ray spectra, the pile-up is too weak to have a significant impact on the work presented here ($<15$\,\% flux loss at the lightcurve peaks). Additionally, since pile-up is roughly proportional to flux, any effect on correlation measurements is minor.
Lightcurves were produced with \textsc{evselect} and \textsc{epiclccorr} and rebinned to match the cadence of the OM frames.

The OM photometry of IRAS~13224--3809 is complicated by a nearby (7.5\,arcsec separation) source (see Fig.~\ref{fig:omimage}) which causes the default execution of \textsc{xmmextractor} to fail.
We therefore take count rates directly from the images using the photometry tool \textsc{imexam} from \textsc{zhtools}, extracting counts from within an aperture of radius 3\,arcsec, using a nearby source-free circular region of radius 18\,arcsec for background subtraction. We correct the count rates for deadtime and coincidence losses using the factors given by \textsc{omichain}. These corrections are between 1.043 and 1.049 for all points apart from one which is 1.029.

In 18 exposures, a count rate less than 0 is returned, which we exclude -- the sky coordinates on these images are wrong (part of OBSID 0792180501). One further point in OBSID 0792180201 is unreasonably low (about 4 times less than neighbouring points) so it is also excluded. This leaves 524 good OM exposures.

We also produce a lightcurve of the nearby source to ensure that it does not affect our results. To minimise the effect of stray light from the edges of the PSF of IRAS~13224--3809, we use a 2\,arcsec radius circular aperture. This shows that the nearby source is too faint and insufficiently variable to affect the lightcurves of IRAS~13224--3809: its flux is 10\,\% of IRAS~13224--3809 and its variability is consistent with Poisson noise.

\subsection{\swift}

\swift\ UVOT \citep{gehrels04,roming05} lightcurves were extracted from level II image files using the tool \textsc{uvotsource}.
We used a circular source region of 5 arcsec radius and a circular background region of 15 arcsec radius from a nearby source free area of the detector.
We excluded exposures where the source region overlaps areas of the detector known to produce low count readings \citep{edelson15}.
The good exposures are then summed across a whole observation.
We converted count rates to fluxes using the conversion factors in \citet{poole08}.
UV fluxes were corrected for Galactic reddening using $E(B-V)=0.0601$ \citep{kalberla05}.
We use lightcurves from all UVOT filters apart from the {\it B}-band, since the three observations available in this band are insufficient to produce reliable variability measurements.

\swift\ XRT \citep{burrows05} lightcurves covering the $0.3-10$\,keV energy band were produced using the online tool available on the UK \swift\ website\footnote{http://www.swift.ac.uk/user\_objects/} \citep{evans07,evans09}. The XRT was operated in PC mode. The source region is a circle of radius 1.2\,arcmin. The background region is an annulus with radii from 2.3 to 7\,arcmin (with point sources removed).

The lightcurves from all instruments are shown in Fig.~\ref{fig:lightcurves}.

Except where noted, errors are given at the 1-$\sigma$ level.

\section{Results}

\begin{figure}
\includegraphics[width=\columnwidth]{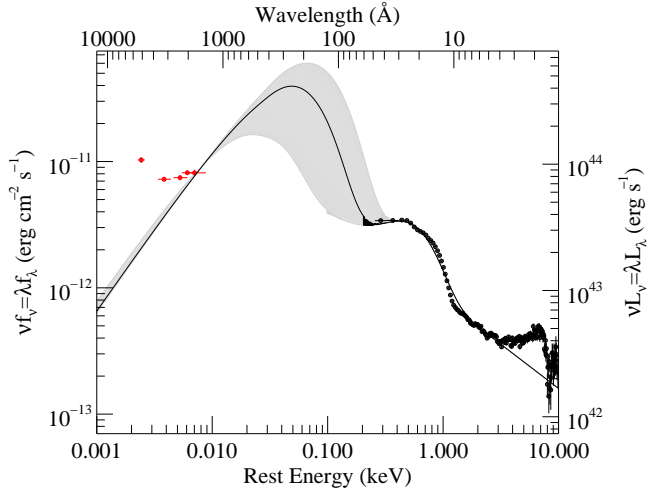}
\caption{
Mean SED of IRAS~13224--3809. Optical/UV points (red) are from \swift-UVOT; X-ray points (black) are from \xmm-pn (Jiang et al., submitted). The grey region indicates the range of SED models used to derive the bolometric luminosity.}
\label{fig:sed}
\end{figure}

\subsection{Mean SED}
\label{sub:sed}

We show the mean SED of IRAS~13224-3809 in Fig.~\ref{fig:sed}. The UV points show the mean flux across the full \swift\ lightcurve; the X-ray points show the mean spectrum from Jiang et al. (submitted). 
We characterise the UV spectrum with a powerlaw of the form $f_\lambda \propto \lambda^\alpha$ and exclude the {\it V}-band since \citet{vandenberk01} show that there is a strong break in powerlaw index at around 5000\,\AA, blueward of the {\it V}-band. This gives $\alpha=-1.2\pm0.1$, slightly softer than the mean quasar spectrum ($\alpha=-1.56$) found in \citet{vandenberk01}, suggesting that there is some contribution from the host galaxy.

From the simultaneous optical to X-ray SED, we can estimate the bolometric luminosity. We approximate the intrinsic AGN emission as a thin disc (for the UV) plus a hot blackbody (for the X-ray soft excess) and a powerlaw (for the hard X-ray component). The principal source of uncertainty is the disc temperature, which is poorly constrained as the cut-off lies in the unobserved extreme UV; we take the lower limit as measured from the RMS spectrum (see Sec.~\ref{sub:uvspec}) and use the X-ray emission to provide the upper limit. This gives a bolometric luminosity range of $4\times10^{44}-1.3\times10^{45}$\,erg\,s$^{-1}$.
For $M_{\rm BH}=10^6-10^7\,{\rm M_\odot}$, this implies an Eddington fraction $\dot{m}=0.3-10$. While this is not a strong constraint (due largely to the poorly-determined black hole mass), a high Eddington fraction is widely regarded as typical of NLS1s and agrees well with estimates of the Eddington fraction from other methods, such as $\dot{m}\simeq0.7$ using the $\Gamma-\dot{m}$ relation of \citet{shemmer08}.

We also consider the relative X-ray and UV power, using the standard measure $\alpha_{\rm OX}$ \citep[e.g.][]{vagnetti10}. This gives $\alpha_{\rm OX}=-1.46$, which is 
compatible with (though at the X-ray weak end of) values found by various authors who have presented a $L_{\rm UV}-\alpha_{\rm OX}$ relation ($\alpha_{\rm OX}=-1.18$, \citealt{gibson08}; $\alpha_{\rm OX}=-1.31$, \citealt{grupe10}; $\alpha_{\rm OX}=-1.46$, \citealt{xu11}).

\begin{figure}
\includegraphics[width=\columnwidth]{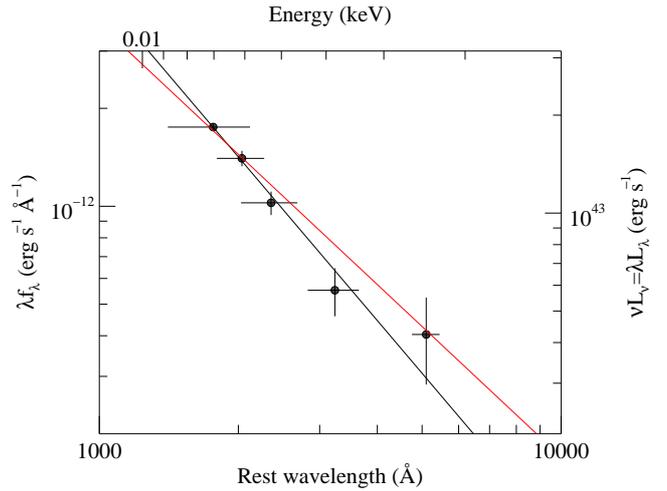}
\caption{
RMS spectrum from {\it Swift}-UVOT data. The black line shows a powerlaw fit, with index $\alpha=-2.67\pm0.15$. The red line has the index expected of a thin disc, $\alpha=-2.33$. Errors in wavelength represent the half maximum of the filters.
}
\label{fig:uvspec}
\end{figure}

\begin{table*}
\caption{Fits to the PSD of the \xmm-OM lightcurve with a powerlaw plus noise model, $P(f)=\alpha (f/10^{-4}\,{\rm Hz})^{\beta} + C$.}
\label{tab:psd}
\centering
\begin{tabular}{lcccc}
\hline
Model & Norm ($\alpha$) & Index ($\beta$) & Noise ($C$) & $\chi^2/{\rm (d.o.f.)}$ \\
\hline
Fixed noise & $ 0.13\pm    0.06$ & $     -1.3\pm     0.3$ & $     0.557$ & $      5.68/
           5$\\
Fixed index & $  0.040\pm    0.013$ & $     -2.0$ & $     0.61\pm    0.05$ & $      8.10/
           5$\\
           Free & $      0.70\pm    0.050$ & $     -0.5\pm     0.1$ & $  <0.35
$ & $       2.45/           4$ \\

\hline
\end{tabular}
\end{table*}

\subsection{\swift\ variable UV spectrum}
\label{sub:uvspec}

To characterise the emission of the innermost regions, we study the variable part of the UV spectrum to avoid contamination by the host galaxy.

We characterise the variable part of the spectrum with the error corrected RMS flux variability, $f_{\rm \lambda,Var}=\sqrt{\sigma^2-\bar{\epsilon}^2}$ \citep{nandra97,edelson02} as in \citet{buisson17}, taking the measured standard deviation, $\sigma$, and mean square error, $\bar{\epsilon}^2$, from the whole lightcurve. Errors on this quantity are given by $ {\rm err}(f_{\rm \lambda,Var}^2)=\frac{1}{\sqrt{N}}\sqrt{\left(\sqrt{2}\bar{\epsilon^2}\right)^2+\left(2\sqrt{\bar{\epsilon^2}}f_{\rm \lambda,Var}\right)^2\bar{x}^2}$ \citep{vaughan03}.

The spectrum this produces (Fig.~\ref{fig:uvspec}) is consistent ($\chi^2/{\rm d.o.f.}=2.8/3$) with a powerlaw, $f_\lambda\propto\lambda^\alpha$, with index $\alpha=-2.67\pm0.15$ (or in frequency units, $f_\nu\propto\nu^\beta$, with $\beta=0.67\pm0.15$). This is consistent at 2-$\sigma$ with the expected index for the emission produced by irradiation of a thin disc \citep[$\alpha=-2$ to --2.33, $\beta=0$ to 0.33,][]{davis07} and significantly flatter than the Rayleigh-Jeans tail of a single-temperature blackbody ($\alpha=-4$, $\beta=2$).

Since the variability is expected to originate in a disc spectrum, we also test a powerlaw with an exponential cut-off representing the maximum temperature at the inner edge of the disc. This places an upper limit of 1440\,\AA\ on the cut-off (at 90\% confidence), corresponding to a blackbody temperature of $\geq10^5$\,K. 
However, such a low cut-off requires a steeper powerlaw index, $\alpha=-4$.
This limit to the temperature is less than that predicted for a standard disc \citep{shakura73}, even for conservative parameters for IRAS~13224--3809 ($M_{\rm BH}=10^{7}\,{\rm M_\odot}$, $\dot{m}=0.1$), so the spectrum is consistent with the temperature of a standard disc.

\subsection{Short timescale optical variability}

\begin{figure}
\includegraphics[width=\columnwidth]{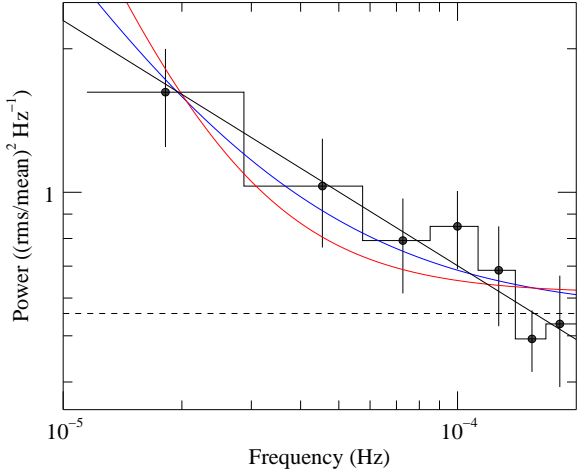}
\caption{PSD of optical monitor data. The estimated Poisson noise level is shown by the dashed line. Solid lines show fits with a powerlaw plus noise model, $P(f)=\alpha f^{\beta} + C$. Red: fixed index ($\beta$), free noise ($C$). Blue: free index, fixed noise. Black: both free. See Table~\ref{tab:psd} for full parameters.}
\label{fig:ompsd}
\end{figure}

While the UVOT lightcurve shows that IRAS~13224--3809 varies over the course of the observing campaign, we also seek to characterise that UV variability on shorter timescales with the \xmm-OM data. 

We calculate the average power spectral density (PSD, \citealt{vaughan03}) of the optical monitor data over the whole observation.
Since calculating the PSD requires an evenly sampled time series, we split the observations where consecutive points are separated by more than 1.5 times the average.
We then take sections of 120\,ks and linearly interpolate onto a regular time grid.
We calculate the periodogram for each section separately and average these into frequency bins containing at least 20 points to give the PSD.
This is shown in Fig.~\ref{fig:ompsd}.
The expected Poisson noise level is calculated from equation A3 in \citet{vaughan03} and shown as the dashed line in Fig.~\ref{fig:ompsd}.
We also fit the PSD with a sum of powerlaw red noise and Poisson white noise. The resulting parameters are shown in Table~\ref{tab:psd}.
The shape of the power spectrum is dependent on the assumptions made about Poisson noise, so we cannot simultaneously constrain the shape of the power spectrum and the level of Poisson noise, which only dominates at higher frequencies.
Owing to the large uncertainties, there is insufficient statistical evidence to choose one model over another.
We expect that fixing the Poisson noise to the calculated value gives the most reliable intrinsic PSD shape, $P(f)\propto f^{1.3\pm0.3}$.
Independent of the exact model chosen, the UV PSD shows that the UV variability has the form of red noise on short timescales.

\subsection{X-ray/UV correlation}

To study the link between the emission from the accretion disc and coronal X-ray emission, we search for correlations between UV and X-ray flux in the \xmm\ observations.

\begin{figure}
\includegraphics[width=\columnwidth]{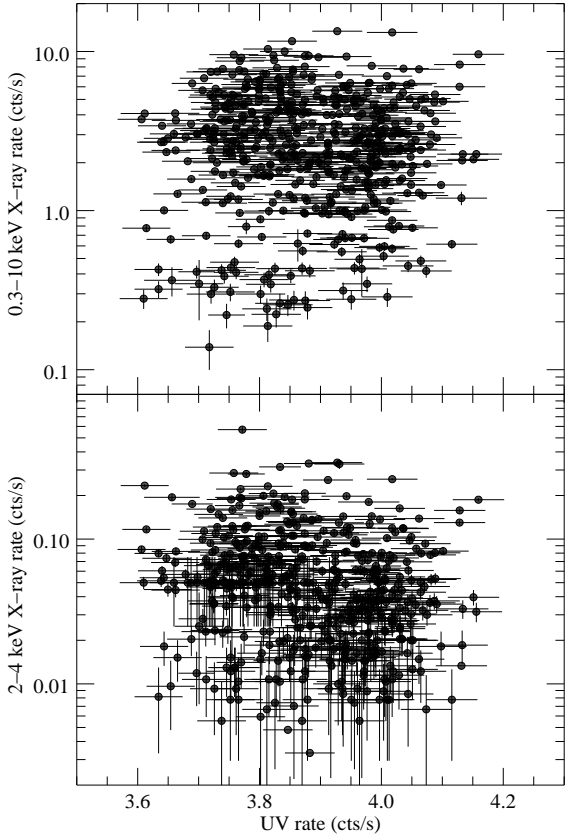}
\caption{Flux-flux plot of X-rays (top: 0.3--10\,keV, full band; bottom: 2--4\,keV, powerlaw dominated) against UV ({\it XMM}-OM {\it W1}, 2910\,\AA). No correlation between the two bands is apparent.}
\label{fig:flfl}
\end{figure}

Initially, we produce a flux-flux plot (Fig.~\ref{fig:flfl}) to detect correlations between simultaneous X-ray and UV emission. This shows no strong correlation between the two bands, with Pearson coefficient $r=-0.02$ ($r=0.025$ in the logarithmic domain) when using the full 0.3--10\,keV band. When drawing lightcurves from uncorrelated red noise (from the same power spectra used for the DCF simulations presented below), a stronger correlation occurs with probability $p=0.95\ (0.94)$. To determine whether the UV correlates with only the primary continuum (rather than the soft excess or reflected emission), we also consider the 2--4\,keV band, which is dominated by the primary emission. This also shows no correlation ($r=-0.15$, $p=0.61$; $r=-0.17$, $p=0.60$ logarithmically) with the UV.

\begin{figure}
\includegraphics[width=\columnwidth]{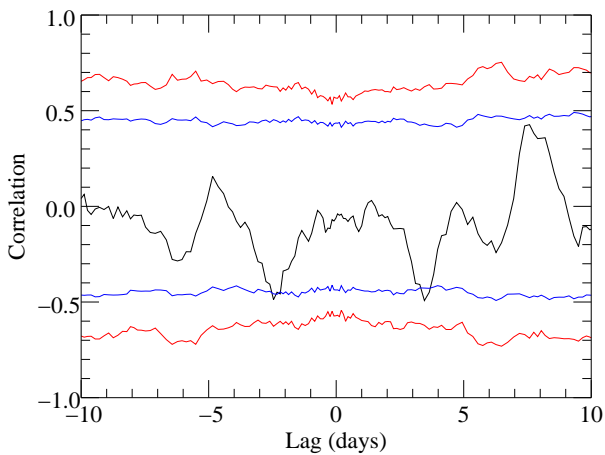}
\caption{DCF of X-rays (0.3--10\,keV) against UV ({\it XMM}-OM {\it W1}, 2910\,\AA). Blue and Red lines indicate 95 and 99\,\% confidence intervals around 0 correlation.}
\label{fig:dcf}
\end{figure}

To test whether the lack of correlation seen in the flux-flux plot is due to a lag between X-ray and UV emission, we use the discrete cross-correlation function \citep[DCF,][]{edelson88} from a single light curve of the whole observation so that timescales up to the full length of the observation are included.

To assess the significance of any correlations, we simulated 10\,000 pairs of uncorrelated light curves and estimated 95 and 99\% confidence intervals from the DCFs measured from these light curves.
We used the method of \citet{timmer95} to generate light curves with appropriate red-noise power spectra. Since the shape of the UV PSD is poorly constrained, we use a simple power law with  $P(f)\propto f^{-\alpha}$ with $\alpha=2$ for the UV. For the X-rays, we use a broken power law with $\alpha=1.1$ and 2.22 below and above $6\times10^{-5}$\,Hz respectively for the X-rays (Alston et al, in prep.).
We extract count rates at times corresponding to the real observations and draw our final simulated data from a Poisson distribution with mean equal to the simulated rates multiplied by the frame time.

The DCF is shown in Fig~\ref{fig:dcf}. There are no significant correlations between the X-ray and optical monitor data. Possible anticorrelations are detected at +3.5 and --2.5 days, although, since there is little physical motivation for such anticorrelations, these may be sampling artefacts due to the gaps between \xmm\ orbits. A spurious detection is not unlikely as 5\% of points are expected to lie outside the 95\% confidence interval.
To test whether only some components of the X-ray emission are correlated with the UV, we test different X-ray bands to isolate the soft excess and powerlaw components; this produces similar results so we show the full band to maximise signal.

\section{Discussion}
\label{section_discussion}

AGN almost universally show variability in their optical to X-ray spectra \citep[e.g.][]{cackett07,ponti12}.
Typically, the UV and X-ray emission is seen to correlate, with the UV often lagging the X-rays, indicative of reprocessing \citep[e.g.][]{edelson15,mchardy16,buisson17}.

We have found that for IRAS~13224--3809, the variability in the UV emission does not clearly correlate with variability observed in X-rays. This lack of correlation is unusual but not unique: for example, 1H~0707--495, which has a similar X-ray spectrum to IRAS~13224--3809 \citep{fabian09}, also shows no correlation between X-ray and UV emission \citep{robertson15}.
While these non-detections use \xmm-OM monitoring covering shorter timescales than are achievable with missions such as \swift, X-ray reprocessing should be detectable in the \xmm\ campaigns: there is strong X-ray variability observed on timescales much shorter than the monitoring campaign.
Indeed, UV/X-ray correlations have been detected with \xmm\ for other sources \citep[e.g.][]{mchardy16} and in shorter \swift\ campaigns \citep[e.g.][]{edelson17,mchardy17,pal17}.
Additionally, AGN cover a wide range of black hole mass and the timescale for variability processes scales linearly with $M_{\rm BH}$. Therefore, timescales probed with \swift\ campaigns for large $M_{\rm BH}$ \citep[e.g. FAIRALL~9, where correlations are observed][]{lohfink14,buisson17} are equivalent to the timescales probed here.
Sources that do not show UV/X-ray correlations must have different emission to typical AGN in one or both of the UV and X-ray bands.

One possibility for the lack of UV/X-ray correlation is that there are significant sources of UV variability other than X-ray irradiation. This is likely to occur in some AGN as \citet{uttley03ngc5548} found more fractional variability in the optical than X-ray emission in NGC~5548 (although in this case the optical and X-ray variability was correlated).
One such source is the intrinsic disc fluctuations which propagate inwards to produce the X-ray variability. However, at the radii which produce the {\it W1}-band emission, the characteristic timescale of these fluctuations is much longer than the observations analysed here.

If the lack of correlation is due to an extra source of UV variability, the UV variability would be expected to be larger than in typical AGN.
On timescales comparable to a night ($2\times10^{-5}-2\times10^{-4}$\,Hz), we find variability of $0.4\pm0.1$\,\%, consistent with
\citet{young99}, who found an upper limit on the optical variability of 1\,\% within a night.
We can also make a direct comparison between the fractional variability of IRAS~13224--3809 and 1H~0707--495. \citet{robertson15} present the fractional variability of 1H~0707--495 in two sets of 4 continuous orbits. To compare the same timescales, we consider the 4 consecutive \xmm\ orbits of IRAS~13224--3809 with OBSIDs 0780561501--0780561801 (other sections of consecutive orbits give similar results).
This epoch has, in the {\it W1}-band, $F_{\rm Var}=1.0\pm0.1$\,\%.
These values are very similar to those of 1H~0707--495, being between the values for the two epochs presented in \citet{robertson15}.

Therefore, both IRAS~13224--3809 and 1H~0707--495 show only modest UV variability, close to the average of 1.2\% found by \citet{smith07} for a sample of AGN measured with the optical monitor. The similarity of the UV variability in both these sources to sample averages \citep[e.g.][]{grupe10} may suggest that it is the nature of their X-ray rather than UV variability which prevents the detection of UV/X-ray correlations.

Despite the lack of UV/X-ray correlation, the variable part of the UV spectrum has the shape expected of an irradiated disc, as found for a number of other AGN in \citet{buisson17} (note that while IRAS~13224--3809 was included in this paper, the \swift\ data at the time of writing were insufficient to produce a RMS spectrum). This also suggests that the lack of UV/X-ray correlation may be due to unusual X-ray rather than UV emission.
To further constrain the nature of the UV/optical emitting region, it would be desirable to study inter-band UV/optical lags, which are sometimes seen to match thin disc expectations even when X-ray lags do not \citep[e.g.][]{edelson17}. However, the available \swift\ data are insufficient to constrain these lags.

\begin{figure}
\includegraphics[width=\columnwidth]{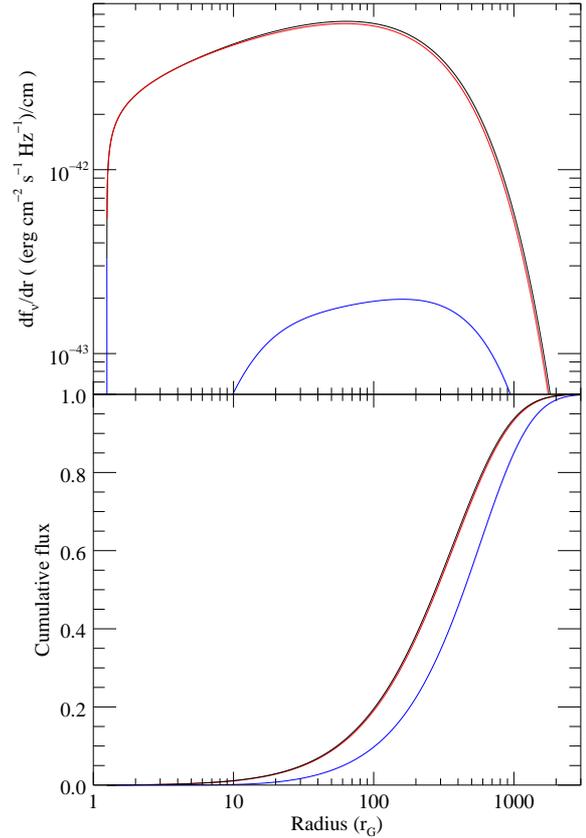}
\caption{Emission at the central wavelength of the {\it W1}-band from a standard disc with representative parameters for IRAS~13224--3809 (see text for details). Black: without X-ray irradiation. Red: with X-ray irradiation. Blue: Difference.}
\label{fig:emis_rad}
\end{figure}

For X-ray heating to be a plausible mechanism to drive the UV variability, there must be sufficient X-ray power to cause the observed changes in UV flux. To determine the regions responsible for {\it W1}-band emission in IRAS~13224--3809, we consider a thin disc \citep{shakura73} illuminated by a central X-ray source \citep{cackett07}.
With sensible parameters for the mass ($M_{\rm BH}=10^7\,{\rm M_{\odot}}$, \citealt{zhou05}) and accretion rate ($\dot{m}=0.7$, \citealt{buisson17}; Jiang et al. submitted), we show the radii responsible for the {\it W1}-band emission in Fig.~\ref{fig:emis_rad}. To demonstrate the potential effect of X-ray irradiation, we also test the same model illuminated by an isotropic point source \citep{cackett07} at $10\,r_{\rm G}$ above the disc, with power $10^{44}$\,erg\,s$^{-1}$ (based on the continuum model in Jiang et al., submitted). This shows that the majority of the flux in the {\it W1}-band is produced on scales of a few hundred $r_{\rm g}$. The change in flux due to heating occurs slightly further out, as more significant flux changes occur when disc material is heated to temperatures at which the material starts to emit in the {\it W1}-band. Integrating the flux density across the disc shows that the X-ray illumination changes the {\it W1}-band flux by $\nu F_\nu({\it W1})=2.5\times10^{-13}$\,erg\,s$^{-1}$. While there is significant uncertainty is some of these parameters, this shows that the effect of X-ray heating can be sufficiently powerful to drive a significant fraction of the observed UV changes.

One alternative model to explain deviations from the simple X-ray reprocessing scenario has been presented by \citet{gardner17}, in which a thickened hot inner disc acts as an intermediate reprocessor between the X-ray and UV emission.
This has been suggested as an explanation for the correlations seen in NGC~5548 \citep{gardner17} and NGC~4151 \citep{edelson17}.
While an additional reprocessor does not remove all correlation between X-ray and UV flux, it significantly reduces the effect of fast X-ray variability on the UV emission. This could mean that X-ray/UV correlations are seen only on long timescales and the campaign presented here is too short to detect a correlation.

The UV variability we do observe could still be due to illumination from the X-ray source if the variability seen by the disc is different to that in our line of sight.
Various effects may lead to different X-ray variability being observed by the disc, such as variable absorption between the disc and corona.
IRAS13224--3809 must have some outflowing material, which may shield the disc from the corona, as a highly ionised variable UFO is observed \citep{parker17nat}. While this outflow is too optically thin to have a significant effect on the transmission of X-ray flux, optically thicker material (denser or less ionised) may exist in the acceleration zone, out of the line of sight, between the corona and disc.
It is also possible for there to be a weak extended region of the corona which, although producing little X-ray flux, is optically thick when viewed from close to the plane of the disc. Scattering in this extended corona could significantly change the flux from the main central corona to the disc relative to that in our line of sight.
If such material is present and changes within the observing campaign (which is seen by \citet{parker17nat} to occur in the highly ionised material) then the X-ray flux which reaches the {\it W1}-band emitting region of the disc may not correlate with the observed X-ray flux.

Alternatively, the changes in X-ray intensity received by the disc may be different to those observed if the geometry of the system changes (such as the corona moving up and down) for several reasons.
Firstly, as the corona rises, it illuminates the disc from a less oblique angle, leading to stronger irradiation of the disc at constant coronal power.
Additionally, if motion of coronal material is at relativistic speeds, changes of this motion will induce differences in the anisotropy of coronal emission due to special relativistic beaming.
General relativistic light bending also acts to focus light towards the black hole \citep{miniutti04,wilkins16}. While this principally affects the innermost regions, small effects in the outer regions may further complicate the observed variability.
A combination of these effects along with changes in the intrinsic coronal power could lead to removal of the correlation between observed coronal power and UV emission from disc heating.
The interpretation of the lack of correlation as being due to variable coronal geometry also fits with the relatively large X-ray variability of IRAS~13224--3809: if other sources have a more stable coronal geometry, they will be observed to have both weaker X-ray variability and stronger X-ray/UV correlation.

This interpretation could be tested with detailed mapping of the corona, such as in \citet{wilkins11,wilkins15mrk335}. This would allow the X-ray irradiation of the disc to be measured rather than just the X-ray flux in the line of sight. However, mapping the corona on sufficiently short timescales is likely to require greater collecting area than is available with current missions.

\section{Conclusions}
\label{section_conclusions}

We have shown that the X-ray and ultraviolet flux of the most X-ray variable bright AGN, IRAS~13224--3809, are not correlated on timescales of up to ${\sim40}$\,days.
However, the variability of the UV spectrum matches that seen in other AGN that do show X-ray/UV correlations.
The UV variability is much weaker than in the X-rays: the average {\it W1}-band fractional variability is $0.7\pm0.1$\% over one \xmm\ orbit and around 3\% over 40\,days, whereas the X-rays vary by more than a factor of ten on timescales of kiloseconds.
This suggests that the X-ray variability viewed by the disc is different to that in our line of sight, which may be caused by changes in coronal geometry, absorption or scattering between the corona and outer disc.

\section*{Acknowledgements}

We thank the referee for comments which have helped to improve the clarity of the paper.
DJKB acknowledges financial support from the Science and Technology Facilities Council (STFC).
ACF, AML and MLP acknowledge support from the ERC Advanced Grant FEEDBACK 340442.
BDM acknowledges support from  the Polish National Science Center grant Polonez 2016/21/P/ST9/04025.
This work has made use of observations obtained with \xmm, an ESA science mission with instruments and contributions directly funded by ESA Member States and NASA.
This work made use of data supplied by the UK Swift Science Data Centre at the University of Leicester.

\bibliographystyle{mnras}
\bibliography{uvpaper}

\bsp	% typesetting comment
\label{lastpage}
\end{document}